\documentclass{emulateapj}
\usepackage{graphicx}
\usepackage{epstopdf}

\newcommand{\Teff}{\mbox{$T_\mathrm{eff}$}}
\newcommand{\Mjup}{\mbox{$M_\mathrm{Jup}$}}
\newcommand{\Msun}{\mbox{$M_{\odot}$}}

\begin{document}
\shorttitle{Near-Infrared Spectroscopy of 2M0441+2301 A\MakeLowercase{ab}B\MakeLowercase{ab}}
\title{Near-Infrared Spectroscopy of 2M0441+2301 A\MakeLowercase{ab}B\MakeLowercase{ab}: \\ A Quadruple System Spanning the Stellar to Planetary Mass Regimes*}

\author{Brendan P. Bowler\altaffilmark{1, 2}, Lynne A. Hillenbrand\altaffilmark{1}
\\ }
\email{bpbowler@caltech.edu}

\altaffiltext{1}{California Institute of Technology, 1200 E. California Blvd., Pasadena, CA 91125, USA.}
\altaffiltext{2}{Caltech Joint Center for Planetary Astronomy Fellow.}
\altaffiltext{*}{The data presented herein were obtained at the W.M. Keck Observatory, which is operated as a scientific partnership 
among the California Institute of Technology, the University of California and the National Aeronautics and Space Administration.  The Observatory was made possible by the generous financial support of the W.M. Keck Foundation.}

\begin{abstract}

We present Keck/NIRC2 and OSIRIS near-infrared imaging and spectroscopy of 2M0441+2301 AabBab, a young (1--3 Myr) hierarchical quadruple system comprising a low-mass star, two brown dwarfs, and a planetary-mass companion in Taurus.  All four components show spectroscopic signs of low surface gravity, and both 2M0441+2301 Aa and Ab possess Pa$\beta$ emission indicating they each harbor accretion subdisks.  Astrometry spanning 2008--2014 reveals orbital motion in both the Aab (0$\farcs$23 separation)  and Bab (0$\farcs$095 separation) pairs, although the implied orbital periods of $>$300 years means dynamical masses will not be possible in the near future.  The faintest component (2M0441+2301 Bb) has an angular $H$-band shape, strong molecular absorption (VO, CO, H$_2$O, and FeH), and shallow alkali lines, confirming its young age, late spectral type (L1 $\pm$ 1), and low temperature ($\approx$1800~K).  With individual masses of 200$^{+100}_{-50}$ \Mjup, 35 $\pm$ 5 \Mjup, 19 $\pm$ 3 \Mjup, and 9.8 $\pm$ 1.8 \Mjup, 
2M0441+2301 AabBab is the lowest-mass quadruple system known.  Its hierarchical orbital architecture and mass ratios imply that it formed from the collapse and fragmentation of a molecular cloud core, demonstrating that planetary-mass companions can originate from a stellar-like pathway analogous to higher-mass quadruple star systems as first speculated by Todorov et al.  More generally, cloud fragmentation may be an important formation pathway for the massive exoplanets that are now regularly being imaged on wide orbits.  

\end{abstract}


\section{Introduction}{\label{sec:intro}}

Over the past two decades a variety of planet-finding techniques have uncovered an unexpectedly diverse array of orbital architectures among extrasolar planetary systems.  Giant planets with masses between 1 to 10 times that of Jupiter have been found spanning over five orders of magnitude in separation from their host stars--- from hot Jupiters at 0.01 AU to planetary-mass companions with projected separations exceeding 1000 AU.  Uncovering the origin and dynamical histories of these populations has emerged as a fundamental goal of planetary science.

Several formation routes have been proposed to explain the disparate orbital properties of giant planets.  In circumstellar disks, the growth of rocky cores and subsequent accretion of gas is generally accepted as the dominant mode of giant planet formation at small separations within about 10 AU (\citealt{Pollack:1996jp}).  On wide orbits of tens of AU to a few hundred AU, gas giants may assemble from pebble accretion (\citealt{Lambrechts:2012gr}) or fragment directly from gravitational instabilities in massive protoplanetary disks during the earliest phases of disk evolution (e.g., \citealt{Boss:1997di}; \citealt{Stamatellos:2009fw}).  Collapsing molecular clouds also naturally give rise to binary systems with companion masses that span low-mass stars, brown dwarfs ($<$75 \Mjup), and possibly opacity-limited fragments with masses below the deuterium-burning limit of about 13 \Mjup \ (\citealt{Low:1976wt}; \citealt{Bate:2009br}; \citealt{Brandt:2014cw}).  All of these mechanisms theoretically can produce gas giants between 1 to 10 Jupiter masses, making it difficult to disentangle the origins of individual planetary systems.  Observational tests of these models are further complicated by high extinction at young ages as well as orbital evolution through disk migration, dynamical scattering from other planets or stars, and secular effects like Kozai-Lidov oscillations, all of which gradually erode the fragile fossilized signatures of planet formation (\citealt{Davies:2013vb}).

\begin{figure*}
  \vskip -.0 in
  \hskip .1 in
  \resizebox{7in}{!}{\includegraphics{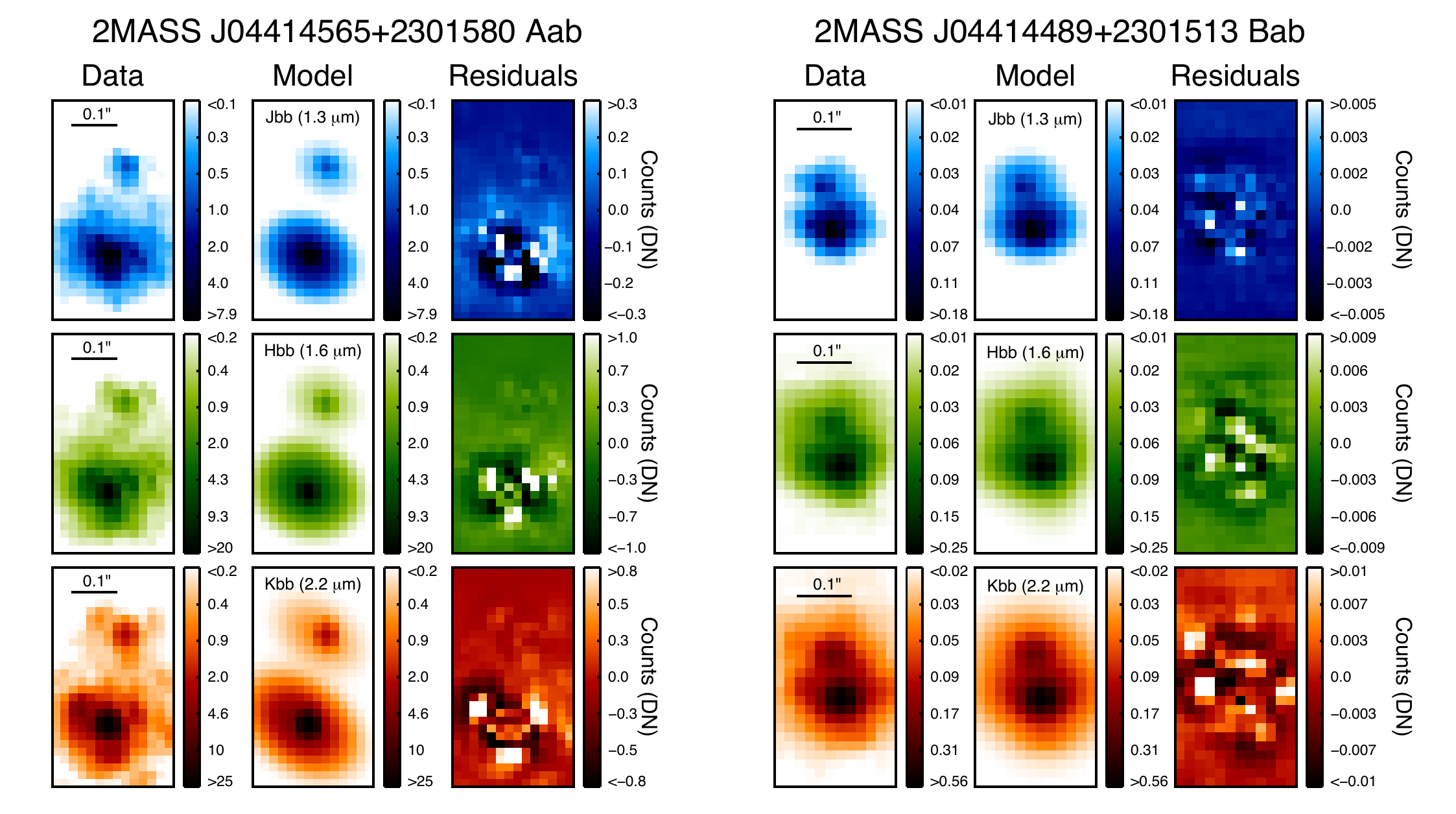}}
  \vskip -0. in
  \caption{Point spread function (PSF) decomposition for extracting spectra from our Keck/OSIRIS observations.  \emph{Left}: Examples of median-combined cubes for 2M0441+2301 Aab in $Jbb$ (top row, blue), $Hbb$ (middle row, green), and $Kbb$ (bottom row, red) filters.  The left panels show the data, the middle panels show our model fits composed of three bivariate Gaussians for each component, and the right panels show the residuals of the fits.  The images are displayed with a logarithmic stretch.  The median rms of the residuals for all observations of the Aab pair are 0.13, 0.23, and 0.31 Data Numbers (DN) for $Jbb$, $Hbb$, and $Kbb$, respectively.  For comparison, the average peak values of the median-collapsed cubes are 10.1, 23.8, and 24.8 DN.  \emph{Right}: Same as the left panels but for 2M0441+2301 Bab.  The rms values for the Bab pair are 0.0056, 0.0097, and 0.012 DN for $Jbb$, $Hbb$, and $Kbb$.  The average peak values are 0.14, 0.27, and 0.51 DN.
 \label{fig:osiris_residuals} } 
\end{figure*}

2M0441+2301AabBab is a young (1--3 Myr) quadruple system in the Taurus star-forming region that offers valuable clues about the ability of cloud fragmentation to form objects below the deuterium-burning limit (\citealt{Todorov:2010cn}).  Indeed, following its discovery based on imaging data from the \emph{Hubble Space Telescope}/WFPC2 and Gemini-N/NIRI, \citet{Todorov:2010cn} suggest this system is most consistent with formation from a fragmented cloud core because of its extremely young age and low mass ratios.  This system consists of two close binaries separated by 12$\farcs$3, or about 1800~AU in projection at the characteristic distance of Taurus (145 $\pm$ 15 parsecs). The more massive pair, 2MASS~J04414565+2301580~Aab (hereinafter 2M0441+2301 Aab), is composed of a low-mass star with an optical spectral type of M4.5 (the ``Aa'' component) orbited by a brown dwarf (``Ab'') at 0$\farcs$23 (33~AU). 2MASS~J04414489+2301513~Bab is composed of an M8.5 brown dwarf (``Ba'') with a fainter companion (``Bb'') at 0$\farcs$1 (15~AU) whose apparent magnitude implies its mass is only about 10 times that of Jupiter.  The hierarchical arrangement of orbits between the compact binaries and much larger binary-binary pair suggests the system is dynamically stable.  Both pairs are well-established members of Taurus (\citealt{Luhman:2009cn}), have little or no reddening ($A_V$ $\sim$ 0.2~mag; \citealt{Bulger:2014ei}), and are confirmed to be physical binaries rather than chance alignments based on multi-epoch astrometry (\citealt{Todorov:2014fq}).  An excess of mid-infrared emission was previously identified from both unresolved pairs, indicating that at least one component of each subsystem harbors a circum-(sub)stellar disk (\citealt{Bulger:2014ei}; \citealt{Adame:2010db}).  Spectroscopic validation is essential to confirm the low temperatures, luminosities, and masses of candidate planets at these very young ages when edge-on disks around brown dwarfs or low-mass stars can cause extinction and mimic the photometric properties of faint protoplanets (\citealt{Kraus:2014tl}; \citealt{Bowler:2014dk}; \citealt{Wu:2015kh}).

Here, we present 1.2--2.4~$\mu$m spatially resolved imaging and spectroscopy of the four components of 2M0441+2301 AabBab.   Our observations are described in Section~\ref{sec:obs}; spectral classification, analysis of emission lines, and atmospheric model fits are detailed in Section \ref{sec:results}; and we discuss the implications of our spectra in the context of planet formation models in Section \ref{sec:discussion}.

\begin{figure*}
  \vskip -.0 in
  \hskip .1 in
  \resizebox{7in}{!}{\includegraphics{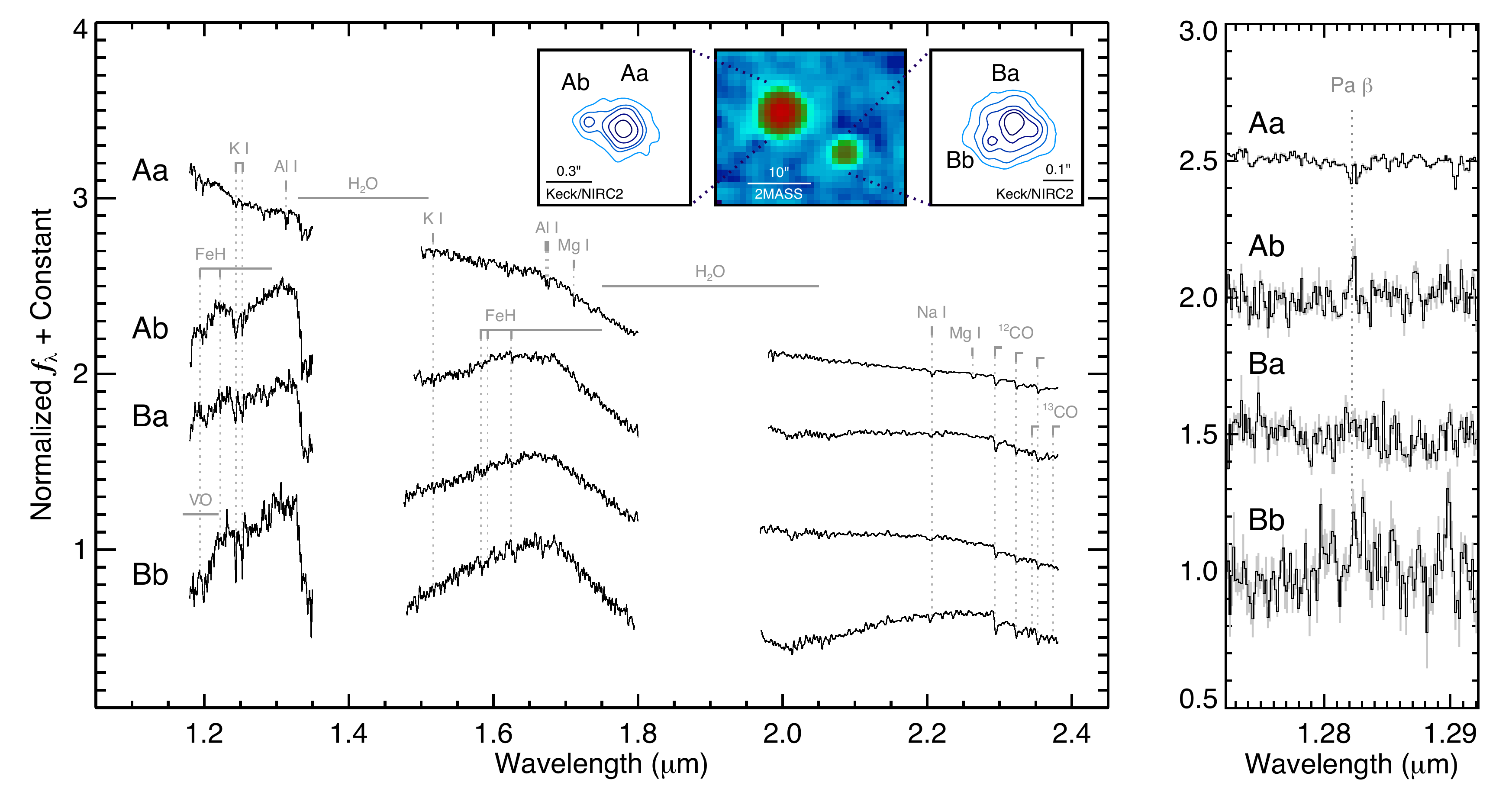}}
  \vskip -0. in
  \caption{Near-infrared spectra of 2M0441+2301 Aa, Ab, Ba, and Bb. \emph{Left}: Flux-calibrated 1.2--2.4 $\mu$m spectra.  Major atomic and molecular features are labeled.  The weak alkali lines such as \ion{K}{1} at 1.25 $\mu$m and triangular $H$-band (1.5--1.8 $\mu$m) shapes in the substellar components are signatures of low surface gravities and extreme youth.  The near-infrared spectral types based on the \citet{Allers:2013hk} classification scheme are M7 $\pm$ 1, M7 $\pm$ 1, and L1 $\pm$ 1 for the Ab, Ba, and Bb components.  All spectra have been smoothed from $R$$\sim$3800 to $R$$\sim$1000 for clarity, normalized, and offset by a constant.  The inset shows the unresolved 2MASS $K_S$-band image of the system together with our AO images of each subsystem in $K_S$ with NIRC2.  Contours for the Aab pair represent 50\%, 20\%, 7\%, 4\%, and 1.5\% of the peak flux.  Contours for the Bab pair represent 50\%, 33\%, 20\%, 10\%, and 5\% of the peak flux.  Note that the Aab image has been convolved with a Gaussian for better rendering. \emph{Right}: The Pa$\beta$ spectral region at the native ($R$$\sim$3800) resolving power.  2M0441+2301 Aa and Ab show weak but significant emission with equivalent widths of --0.96 $\pm$ 0.02 \AA \ and --0.84 $\pm$ 0.11 \AA, respectively.  The gray shaded region shows the 1 $\sigma$ flux density uncertainties.
 \label{fig:specfig} } 
\end{figure*}

\section{Observations}{\label{sec:obs}}

We obtained moderate-resolution ($\lambda$/$\Delta$$\lambda$ $\sim$ 3800) integral field spectroscopy of 2M0441+2301 Aab and Bab with the OH-Suppressing Infrared Imaging Spectrograph (OSIRIS; \citealt{Larkin:2006jd}) coupled with laser guide star adaptive optics (AO; \citealt{Wizinowich:2006hk}; \citealt{vanDam:2006hw}) at the Keck I telescope on 2014 December 08 Universal Time (UT).  OSIRIS samples over 1000 spectra in a single observation, resulting in a three-dimensional data cube with a field of view of 0$\farcs$32 $\times$ 1$\farcs$28 spanning about 1600 spectral channels for the 0$\farcs$020/pixel plate scale with broad band filters.  Our observations benefited from the recently upgraded ``G3'' grating (\citealt{Mieda:2014dt}).  

2M0441+2301 Aa ($R$ = 14.2~mag) was used as a reference tip tilt star for both binaries.  The $R$-band magnitude of the laser guide star on the wavefront sensor remained steady at 9.3~mag during all the observations.  Sky conditions were clear with excellent seeing between 0.4--0.6$''$ in the optical as recorded by the Differential Image Motion Monitor at the Canada-France-Hawaii Telescope.  We obtained a total of 40 min of on-source integration for 2M0441+2301 Aab  in $Jbb$ band (1.18--1.42 $\mu$m), 30 min in $Hbb$ band (1.47--1.80~$\mu$m), and 30 min in $Kbb$ band (1.97--2.38~$\mu$m) over an airmass of 1.0 to 1.2.  For 2M0441+2301 Bab, we obtained 67 min in $Jbb$, 40 min in $Hbb$, and 40 min in $Kbb$ spanning airmasses between 1.0--1.6 (Table~\ref{tab:obs}).  The instrument rotator was set to 90$^{\circ}$ for 2M0441+2301 Aab and 120$^{\circ}$ for Bab, and the observations were executed in an ABBA pattern with 0.4--0.5$''$ dithers along the long axis of the reconstructed data cube.  The A0V standard stars HD~19600 and HD~35036 were targeted immediately before and after both science objects with airmasses well matched to the associated science frames for each bandpass. 

Basic image reduction including bad pixel removal, flat fielding, pair-wise subtraction, wavelength calibration (in vacuum wavelengths), and assembly of the 2-D spectra into 3-D cubes was carried out using version 3.2 of the OSIRIS Data Reduction Pipeline with the most recent rectification matrices maintained by Keck Observatory.  The current separation between 2M0441+2301Ba and Bb is 95 mas, which is only 1.8 diffraction limits at the 10-meter Keck telescope in $K$ band. The fainter companions in each binary pair (Ab and Bb) are easily visible in all median-collapsed data cubes, but their PSFs overlap with those of their primaries, so robust spectral extraction free of systematics is not trivial.  Our solution is to simultaneously fit analytical PSF models to each binary component by making use of the nonlinear least-squares curve fitting routine \texttt{MPFIT} (\citealt{Markwardt:2009wq}) written in Interactive Data Language (IDL).  We tested four models for each binary component: a single bivariate Gaussian, a single Lorentzian, the sum of a bivariate Gaussian and Lorentzian, and the sum of three bivariate Gaussians.  These increasingly complex models are designed to mimic the extended PSF halo and a more compact diffraction-limited core.  

The PSFs of each component in our OSIRIS data have the same shape, remain at fixed relative positions in each cube, and both broaden in the same fashion at longer wavelengths, but their amplitudes vary with wavelength because of differing spectral slopes and absorption features.  We therefore first fit each joint model to the binary in the median-combined cube to establish the position of each model component, which is then held fixed for the PSF fit at each wavelength.  All other parameters defining the PSF are allowed to vary, but the shape of each component (Aa and Ab, for example) is designed to remain mutually identical.   Ultimately, we adopted the model composed of three Gaussians per binary component (i.e., six Gaussians per binary) because this produced the best fit as measured by the lowest residual rms.  Fourteen free parameters define the Gaussian amplitudes, standard deviations, position angles, and overall offset of the binary model (Figure~\ref{fig:osiris_residuals}) and was used to fit our observations at each wavelength channel of every data cube and in all three filters for both 2M0441+2301 Aab and Bab.  The same model (except for a single component instead of a binary) was used to extract the spectrum of the standard stars.

The extracted spectra are scaled to a common level and the mean and standard error at each wavelength are adopted as the flux density and its associated uncertainty.  Telluric correction was carried out with the \texttt{xtellcor\_general} routine in the IRTF/SpeX reduction package Spextool, which is written in IDL (\citealt{Vacca:2003wi}; \citealt{Cushing:2004bq}).  Each band was independently flux calibrated by first converting unresolved 2MASS photometry of each subsystem into the OSIRIS filter system by deriving photometric transformations as a function of spectral type from synthetic photometry of MLT objects from the SpeX Prism Spectral Library (\citealt{Burgasser:2014tr}).  Binary flux ratios are measured during the modeling and spectral extraction of the science data and are used to calculate apparent magnitudes of each component in the OSIRIS filter system.  Finally, a scale factor is derived to absolutely flux calibrate each bandpass of the spectrum to match the decomposed OSIRIS photometry (Figure~\ref{fig:specfig}).  The signal-to-noise ratio per pixel is lower near the edges of each bandpass and is highest in the centers with median values across the entire 1.2--2.4$\mu$m spectrum of 112 for Aa, 45 for Ab, 41 for Ba, and 24 for Bb.

In addition to our OSIRIS observations, we also obtained NIRC2 images of 2M0441+2301 AabBab at the Keck II telescope on 2014 November 10 UT using Natural Guide Star (NGS) AO.  Sky conditions were clear with 0.5--0.6$''$ seeing in the optical throughout the observations.  We targeted the Aab component in $Y$, $J$ (on the Mauna Kea Observatory, or MKO, filter system), $H$ (MKO), and $K_S$ filters and the Bab component in $H$ (MKO), and $K_S$ filters with the narrow camera mode, resulting in a field of view of 10$\farcs$2$\times$10$\farcs$2.  A plate scale of 9.952 $\pm$ 0.002 mas pix$^{-1}$ and sky orientation on the detector of 0.252 $\pm$ 0.009$^{\circ}$ are adopted from \citet{Yelda:2010ig}.  2M0441+2301 Aa provided on-axis AO correction for the Aab pair, and we offset by 12$\farcs$3 for off-axis imaging of the Bab components.  The images were dark subtracted, flat fielded, and cleaned of cosmic rays and bad pixels.  Relative photometry and astrometry was measured by fitting three bivariate Gaussians to each binary component in each image following \citet{Liu:2010cw}.  We adopt the mean and standard deviation of these measurements, which are reported in Table~\ref{tab:obs}.

\begin{figure}
  \vskip -.2 in
  \hskip -.2 in
  \resizebox{3.9in}{!}{\includegraphics{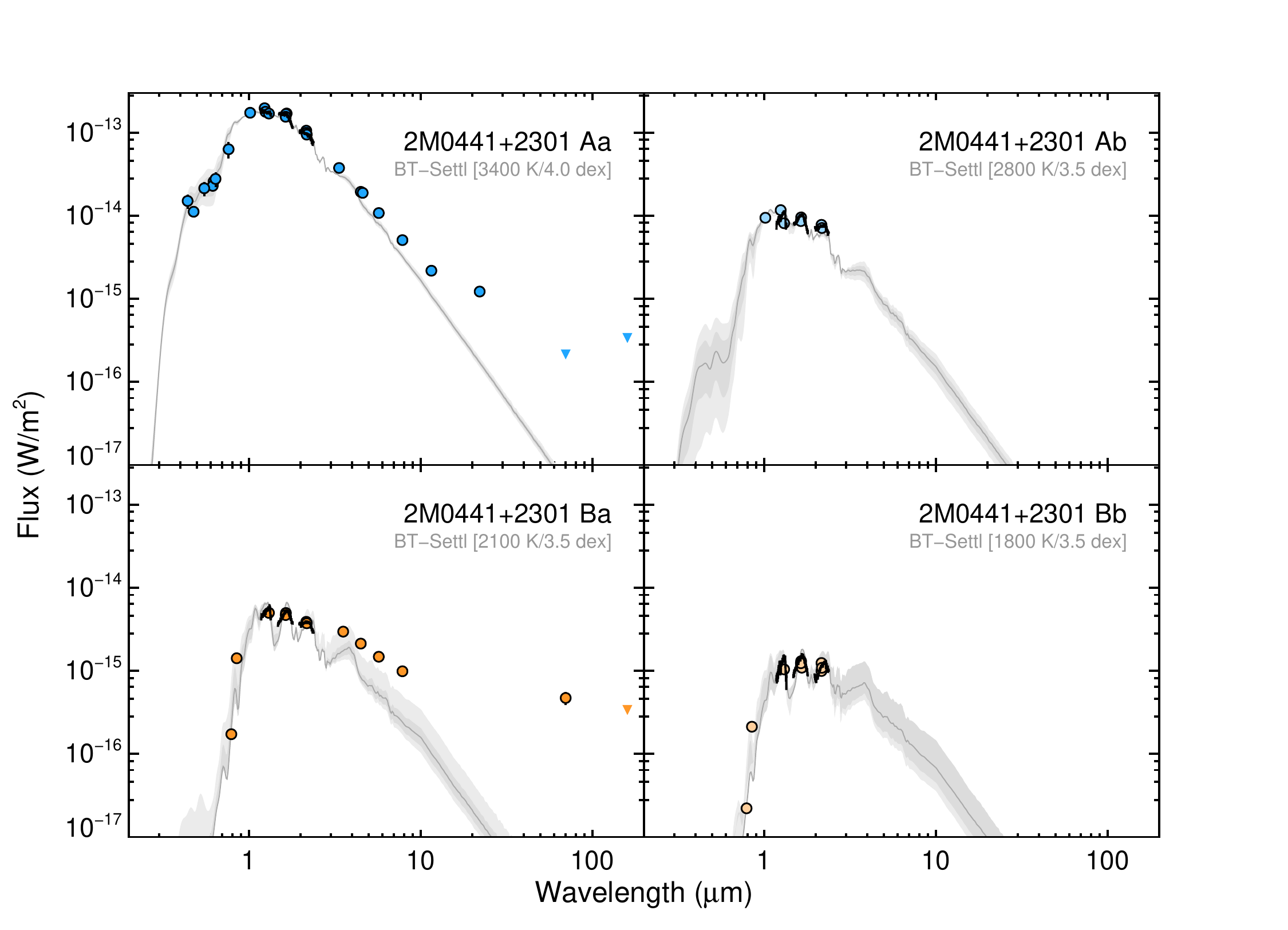}}
  \vskip -.1 in
  \caption{Spectral energy distributions for 2M0441+2301 AabBab.  Dark gray lines show the best matching BT-Settl models (as indicated in the legend) to $<$3 $\mu$m photometry assuming $A_V$=0.0~mag, which avoids disk emission at longer wavelengths.  Shaded regions show the 1 $\sigma$ and 2 $\sigma$ uncertainties.  Photometry originates from the literature (\citealt{Todorov:2010cn}; \citealt{Bulger:2014ei}) and our own NIRC2 and OSIRIS observations.  Unresolved excess emission from one or more subdisks is present beyond about 7 $\mu$m for the Aab component and 3 $\mu$m for Bab, shown here as associated with the primaries of each pair but which may also (or instead) originate from the companions.  The flux-calibrated OSIRIS spectra are overplotted in black for comparison.
 \label{fig:sedfig} } 
\end{figure}

\begin{figure*}
  \vskip -.1 in
  \hskip .2 in
  \resizebox{6.4in}{!}{\includegraphics{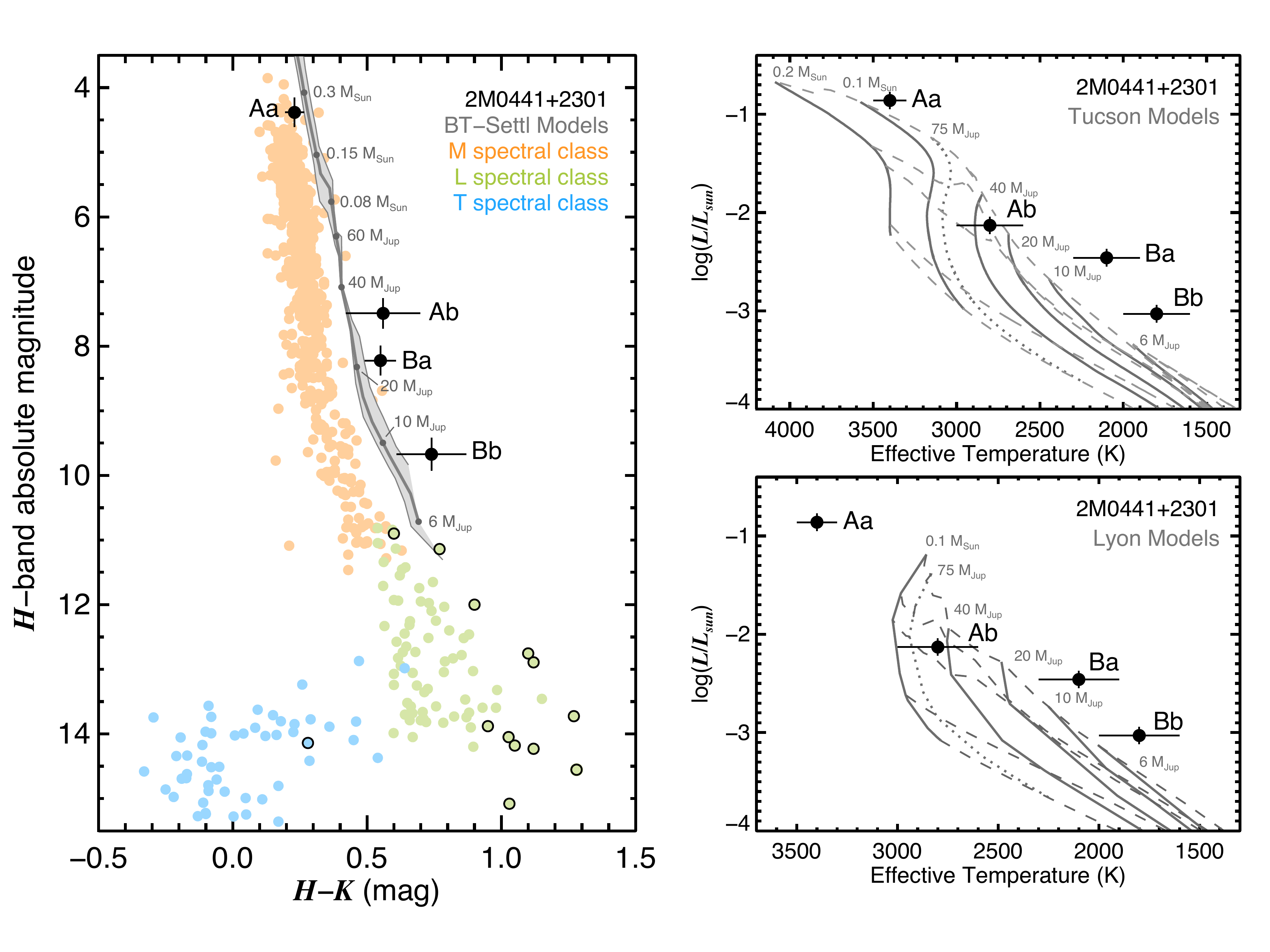}}
  \vskip -.1 in
  \caption{Comparison of 2M0441+2301 AabBab with theoretical isochrones.  \emph{Left}: The BT-Settl models (\citealt{Baraffe:2015fw}) spanning 1, 2, and 3 Myr isochrones from right to left are consistent within 1 $\sigma$ with 2M0441+2301 AabBab in the $M_H$ versus $H$--$K$ near-infrared color-magnitude diagram.  All photometry has been converted to the MKO filter system.  Comparison low-mass field objects are from the literature (\citealt{Dupuy:2012bp}), and symbols with dark borders represent young brown dwarfs and giant planets as opposed to old stars and brown dwarfs.  \emph{Top Right}: The 1 Myr isochrones from the Tucson models (\citealt{Burrows:1997jq}) are generally consistent with the empirical isochrone traced by this quadruple system at the 2 $\sigma$ level, though the Ba and Bb components are cooler than expected from these models by 200--300~K.  1 Myr, 5 Myr, 10 Myr, 100 Myr, and 1 Gyr isochrones are plotted.  \emph{Bottom Right}: Same as the top right panel, but for the Lyon (Cond) evolutionary models (\citealt{Baraffe:2003bj}).  These models extend to 0.1 \Msun \ and are consistent with 2M0441+2301 Ab, Ba, and Bb at the 2 $\sigma$ level. 1 Myr, 5 Myr, 10 Myr, 100 Myr, and 1 Gyr isochrones are plotted. 
 \label{fig:cmd} } 
\end{figure*}

\section{Results}{\label{sec:results}}

\subsection{Spectral Classification}

The flux-calibrated spectra of 2M0441+2301 AabBab are shown in Figure~\ref{fig:specfig}.  The substellar components Ab, Ba, and Bb show strong atomic and molecular absorption features characteristic of young late-M to early-L type brown dwarfs with low surface gravities.  These include shallow \ion{K}{1} absorption lines at 1.25~$\mu$m and an angular spectral shape from 1.5 to 1.8 $\mu$m compared to old field objects (\citealt{Allers:2013hk}), which is primarily caused by weakened collision-induced absorption of molecular hydrogen at lower photospheric pressures (\citealt{Marley:2012fo}; \citealt{Barman:2011fe}).  2M0441+2301 Bb shows particularly pronounced features from water, carbon monoxide, iron hydride, and vanadium oxide.  Index-based classification of the substellar components (\citealt{Allers:2013hk}) gives near-infrared spectral types of M7 $\pm$ 1, M7 $\pm$ 1, and L1 $\pm$ 1 for 2M0441+2301 Ab, Ba, and Bb.  We measure 1.2437 $\mu$m and 1.2529 $\mu$m \ion{K}{1}  equivalent widths of 3.5 $\pm$ 0.3~\AA \ and 0.0 $\pm$ 0.3~\AA \ for Ab; 2.3 $\pm$ 0.3~\AA \ and 0.8 $\pm$ 0.5~\AA \ for Ba; and 3.4 $\pm$ 0.6~\AA \ and 5.5 $\pm$ 1.0~\AA \ for Bb, respectively.  Aa and Ba have very low gravity (``VL-G'') classifications, while Bb has an intermediate-gravity classification (``INT-G'').

\subsection{Emission Lines and Accretion}

2M0441+2301 Aa and Ab show weak Pa$\beta$ emission indicating ongoing accretion from subdisks around both binary components.  No significant Pa$\beta$ emission is evident in 2M0441+2301 Ba or Bb, and there is no Br$\gamma$ emission in any components.  The emission line from the primary is superimposed on a slightly broader absorption trough from unresolved \ion{Ti}{1}, \ion{Fe}{1}, and \ion{Ca}{1} absorption lines (\citealt{Rayner:2009ki}).  We therefore define a pseudocontinuum level at 1.25 $\pm$ 0.01 $\times$ 10$^{-13}$ W m$^{-2}$  $\mu$m$^{-1}$ to compute the equivalent width of the emission line (--0.96 $\pm$ 0.02 \AA), line flux (1.20 $\pm$ 0.04 x10$^{-17}$ W m$^{-2}$), and line luminosity (log($L_\mathrm{Pa \beta}$/$L_{\odot}$) = --5.11 $\pm$ 0.09~dex).  The corresponding accretion luminosity is log($L_\mathrm{acc}$/$L_{\odot}$) = --2.9 $\pm$ 1.0 dex using the empirical correlation between accretion luminosity and Pa$\beta$ line luminosity from \citet{Natta:2004kz}.  We find a mass accretion rate of log(\emph{\.{M}}) = --9.4 $\pm$ 1.1 \Msun \ yr$^{-1}$ using \emph{\.{M}} = 2$L_\mathrm{acc}$$R_*$/($GM_*$), where $G$ is the gravitational constant and $M_*$ is the stellar mass.  We adopt a stellar radius, $R_*$, of 1.1 $\pm$ 0.2 $R_{\odot}$ based on the age and luminosity of the system and evolutionary models (\citealt{Allard:2012fp}) and a stellar mass of 0.20$^{+1.0}_{-0.05}$ $M_{\odot}$ (Section \ref{sec:masses}).

For 2M0441+2301~Ab, we use the $\pm$ 0.01 $\mu$m region adjacent to the Pa$\beta$ line to define the pseudocontinuum level and measure a Pa$\beta$ equivalent width of --0.84 $\pm$ 0.11 \AA.  This corresponds to a line flux of 6.7 $\pm$ 0.9 $\times$ 10$^{-19}$ W m$^{-2}$, a line luminosity of log($L_\mathrm{Pa \beta}$/$L_{\odot}$) = --6.4 $\pm$ 0.1 dex, an accretion luminosity of log($L_\mathrm{acc}$/$L_{\odot}$) = --4.7 $\pm$ 1.3 dex, and a mass accretion rate of log(\emph{\.{M}}) = --10.8 $\pm$ 1.3 \Msun \ yr$^{-1}$ assuming a radius of 0.41 $\pm$ 0.02 $R_*$ (\citealt{Baraffe:2003bj}) and a mass of 35 $\pm$ 5 \Mjup \ (Section \ref{sec:masses}).  

\subsection{Atmospheric Model Fitting}

Atmospheric models can be used to infer the physical properties of the 2M0441+2301AabBab system such as the components' effective temperatures (\Teff) and surface gravities (log $g$) by fitting synthetic spectra to our OSIRIS observations.  We first fit the BT-Settl models (\citealt{Allard:2012fp}) with ``CIFIST2011'' solar abundances (\citealt{Caffau:2010ik}) to individual $J$, $H$, and $K$ spectral regions and to the entire flux-calibrated 1.2--2.4 $\mu$m spectrum of all components in the system.  The grid spans 1100~K to 4000~K in \Teff ($\Delta$\Teff = 100~K) and 2.5~dex to 5.5~dex in log $g$ ($\Delta$log $g$ = 0.5 dex [cgs units]).  For each synthetic spectrum, we compute the scale factor that minimizes the $\chi^2$ value between the model and the data.  This also gives the objectÕs radius $R$ since the scale factor is simply $R^2$/$d^2$, where $d$ is the distance to the object.  We adopt the characteristic distance to Taurus of 145 $\pm$ 15 pc throughout this work.  The models qualitatively reproduce our near-infrared spectra with various combinations of temperature and gravity, but the best-fit values to different bandpasses are generally not self-consistent and are therefore unreliable, yielding unphysical values that differ by up to 1300~K and 2.0~dex for the same object.  This is likely a result of missing opacity sources, incomplete atomic and molecular line lists, and imperfect cloud prescriptions in the models.  For example, although fits to the individual bandpasses of 2M0441+2301~Bb provide good reduced chi-squared values, the model effective temperatures range from 1600~K to 2500~K and the inferred radii span 1.6 to 5.4 $R_\mathrm{Jup}$.  

An alternative approach is to use photometry to constrain the components' effective temperatures since this spans a broader range of wavelengths and is less sensitive to the detailed physics and input line lists compared to our near-infrared spectra.  After flux calibrating the models to the $H$-band monochromatic flux density, we identify the best matches to the available photometry of all four objects spanning 0.4--3 $\mu$m.  As with comparisons to our spectra, the best-fitting models often have unphysical surface gravities.  We therefore fix the surface gravities to the predicted values from the \citet{Baraffe:2003bj} evolutionary models because gravity is only minimally influenced by model assumptions.  We adopt values of log $g$ = 4.0~dex for Aa and log $g$ = 3.5~dex for Ab, Ba, and Bb and compute $\chi^2$ values for a range of effective temperatures (1500 K to 4000 K).  The best matches have effective temperatures of 3400 K for Aa, 2800 K for Ab, 2100 K for Ba, and 1800 K for Bb.  

The results for Ba and Bb are up to 500 K cooler than recent spectral type-effective temperature calibrations for young stars from \citet{Herczeg:2014is}, which are derived by fitting the same grid of BT-Settl atmospheric models we use in this work to optical spectra of T Tauri stars.  This discrepancy may reflect the paucity of photometric observations for 2M0441+2301 Ab, Ba, and Bb shortward of $J$ band where the SED turns over and provides the best constraint on \Teff.  The reduced $\chi^2$ values ($\chi^2_{\nu}$) are much larger than unity ($\chi^2_{\nu}$ $\sim$ 10 to 300), implying the photometric uncertainties are underestimated and/or there are systematic errors in the models.  Since these data do not appear to be standard deviates of the models, maximum likelihood methods for establishing parameter confidence regions like constant $\Delta$$\chi^2$ boundaries about the minimum $\chi^2$ values are not valid here.  We therefore adopt a single grid step (100~K) as an estimate of the uncertainty in \Teff \ for Aa and two grid steps (200~K) for the substellar objects.  Note that we have assumed zero reddening for these fits, but reddening up to $A_V$ = 0.5~mag does not change the fitting results by more than 1 $\sigma$.  Additional resolved optical photometry would provide better constraints on \Teff \ and log $g$, which in turn would enable more rigorous tests of substellar evolutionary models.

\subsection{Luminosities and Masses}{\label{sec:masses}}

Integrating the models and adopting the mean Taurus distance gives luminosities of log($L$/$L_{\odot}$) = --0.86 $\pm$ 0.09 dex, --2.13 $\pm$ 0.09 dex, --2.46 $\pm$ 0.09 dex, and --3.03 $\pm$ 0.09 dex for Aa, Ab, Ba, and Bb, respectively.  Assuming a uniform likelihood of ages between 1 to 3 Myr, the corresponding masses of the substellar components are 35 $\pm$ 5 \Mjup \ for Ab, 19 $\pm$ 3 \Mjup \ for Ba, and 9.8 $\pm$ 1.8 \Mjup \ for Bb from interpolated substellar evolutionary models (\citealt{Baraffe:2003bj}). We adopt a mass of 0.20$^{+1.0}_{-0.05}$ $M_{\odot}$ for the primary star Aa based on recent low mass stellar evolutionary models (\citealt{Baraffe:2015fw}).  The mass we find for Bb is somewhat greater than \citet{Todorov:2010cn} owing to our slightly higher  luminosity measurement and broader assumed age range of 1 to 3 Myr.  

Altogether this confirms that 2M0441+2301 Bb is similar in mass and separation to directly imaged planets like HR 8799 bcde (\citealt{Marois:2008ei}; \citealt{Marois:2010gpa}) and $\beta$~Pic~b (\citealt{Lagrange:2010fsa}).  The mass ratios of these systems, however, differ by nearly two orders of magnitude; the HR 8799 and $\beta$ Pic systems have mass ratios of about 150--220, but for the 2M0441+2301 Bab binary it is only about 1.9.  Together with environmental clues such as apparently coplanar multiple planets and an unusually massive and extended debris disk, the high mass ratios of HR 8799 and $\beta$ Pic point to an alternative formation route compared with 2M0441+2301 Bab (see Section \ref{sec:discussion}). Finally, these results also confirm that 2M0441+2301 AabBab is the lowest-mass quadruple system currently known, with a total mass of about 0.26 $M_{\odot}$. 

\section{Discussion}{\label{sec:discussion}}
 
As a coeval and presumably chemical homogeneous quadruple system spanning the stellar to planetary mass regimes, 2M0441+2301 AabBab also provides a unique opportunity to test pre-main sequence and giant planet evolutionary models similar to other high-order stellar/substellar multiple systems in Taurus like GG Tau (\citealt{White:1999bt}).   Figure \ref{fig:cmd} shows the location of 2M0441+2301~AabBab in color-magnitude and Hertzsprung-Russell diagrams.  Evolutionary model isochrones spanning 1 to 3 Myr are broadly consistent with the data at the 1 to 2 $\sigma$ levels.
  
We detect orbital motion for the first time in both the Aab and Bab binaries.  2M0441+2301Aab has undergone significant changes in separation of 1.4 $\pm$ 0.2 mas yr$^{-1}$ and in position angle (P.A.) of --1.3 $\pm$ 0.02$^{\circ}$ yr$^{-1}$ since 2008 (\citealt{Kraus:2011hv}; \citealt{Todorov:2010cn}; \citealt{Todorov:2014fq}).  2M0441+2301~Bab is moving by --1.5 $\pm$ 0.5 mas yr$^{-1}$ in separation and 2.0 $\pm$ 0.3$^{\circ}$ yr$^{-1}$ in P.A.  Unfortunately, the long orbital periods of about 300--400 years for each subsystem means that dynamical mass measurements will not be possible for at least several decades.

The wide separation between 2M0441+2301~Aab and Bab, the hierarchical orbital architecture of the system, and the low mass ratios of the Ab, Ba, and Bb components relative to the primary star Aa argue against formation by core accretion or stellar capture.  Simulations of fragmenting disks can result in close substellar companions and ejected brown dwarf-planetary mass object binaries, but such low mass quadruple systems with a near ejection of one pair is difficult to reproduce with this formation channel (\citealt{Li:2015ib}).  Instead, the 2+2 hierarchy of 2M0441+2301 AabBab resembles the configuration of the most common stellar quadruples, which outnumber 3+1 hierarchies in Òplanetary-likeÓ configurations by at least 4 to 1 among Sun-like stars within 25 pc (\citealt{Tokovinin:2014ea}).  High order multiple systems like 2M0441+2301 AabBab provide important clues about the masses, critical densities, characteristic sizes, and opacity limits of fragmenting molecular cloud cores.  This is particularly true at young ages and in low density star forming regions before significant dynamical unfolding has occurred and where interactions with nearby stars are minimal (\citealt{Reipurth:2015dz}; \citealt{Pineda:2015iz}).  Although the details and universality of quadruple star formation remain unclear, one possible formation route for this system is through two primary fragmentation events which each subsequently underwent a secondary fragmentation following H$_2$ dissociation, resulting in two short-period binaries on weakly bound orbits about each other (\citealt{Whitworth:2006cd}).  If the current separation of at least 1800~AU reflects the nascent architecture of the system, then the binary pairs have undergone fewer than 20 full orbits around each other in 3~Myr.

The lower limit of cloud fragmentation has been extensively explored with increasingly sophisticated simulations of collapsing turbulent molecular clouds (\citealt{Bate:2009br}).  Opacity-limited fragments between about 3 to 10 \Mjup \ can be produced if accretion is rapidly halted through dynamical encounters or ejections out of dense cloud cores (\citealt{Padoan:2004ez}).  This is generally consistent with empirical determinations of the initial mass function, which show a smooth and continuous distribution down to 5 to 10 \Mjup \ and suggest a common origin with most stars and brown dwarfs (\citealt{Chabrier:2003ki}).  This mass range also coincides with the lowest-mass companions imaged around stars and brown dwarfs (\citealt{Brandt:2014cw}).  However, the formation pathway of individual systems is often ambiguous since free-floating planetary-mass objects could represent ejections from planet-planet scattering events, and directly imaged planets could be scattered or in situ products of disk instability or core accretion.  Confirming the low mass of 2M0441+2301 Bb demonstrates that fragmenting molecular clouds can create companions which overlap in mass with bona fide planets formed in disks, implying that some of the directly imaged planets orbiting stars and brown dwarfs may represent the tail end of turbulent fragmentation rather than the tip of the iceberg of planet formation.

\acknowledgments

We thank K. Deck for helpful suggestions for improving this article, M. Brown for contributing telescope time for our NIRC2 observations, and the entire Keck Observatory staff for their exceptional support.  We utilized data products from the Two Micron All Sky Survey, which is a joint project of the University of Massachusetts and the Infrared Processing and Analysis Center/California Institute of Technology, funded by the National Aeronautics and Space Administration and the National Science Foundation.  NASA's Astrophysics Data System Bibliographic Services together with the VizieR catalogue access tool and SIMBAD database operated at CDS, Strasbourg, France, were invaluable resources for this work.  This research has benefitted from the SpeX Prism Spectral Libraries, maintained by Adam Burgasser at http://pono.ucsd.edu/$\sim$adam/browndwarfs/spexprism.  Finally, mahalo nui loa to the kama`\={a}ina of Hawai`i for their support of Keck and the Mauna Kea observatories.  We are grateful to conduct observations from this mountain.

\facility{{\it Facilities}: \facility{Keck:II (NIRC2)}, \facility{Keck:I (OSIRIS)}}



\begin{thebibliography}{}
\expandafter\ifx\csname natexlab\endcsname\relax\def\natexlab#1{#1}\fi

\bibitem[{Adame {et~al.}(2010)Adame, Calvet, Luhman, D'Alessio, Furlan,
  McClure, Hartmann, Forrest, \& Watson}]{Adame:2010db}
Adame, L., Calvet, N., Luhman, K.~L., {et~al.} 2010, The Astrophysical Journal,
  726, L3

\bibitem[{Allard {et~al.}(2012)Allard, Homeier, \& Freytag}]{Allard:2012fp}
Allard, F., Homeier, D., \& Freytag, B. 2012, Philosophical Transactions of the
  Royal Society A: Mathematical, Physical and Engineering Sciences, 370, 2765

\bibitem[{Allers \& Liu(2013)}]{Allers:2013hk}
Allers, K.~N., \& Liu, M.~C. 2013, The Astrophysical Journal, 772, 79

\bibitem[{Baraffe {et~al.}(2003)Baraffe, Chabrier, Barman, Allard, \&
  Hauschildt}]{Baraffe:2003bj}
Baraffe, I., Chabrier, G., Barman, T.~S., Allard, F., \& Hauschildt, P.~H.
  2003, A{\&}A, 402, 701

\bibitem[{Baraffe {et~al.}(2015)Baraffe, Homeier, Allard, \&
  Chabrier}]{Baraffe:2015fw}
Baraffe, I., Homeier, D., Allard, F., \& Chabrier, G. 2015, arXiv, 1503.04107v1

\bibitem[{Barman {et~al.}(2011)Barman, Macintosh, Konopacky, \&
  Marois}]{Barman:2011fe}
Barman, T.~S., Macintosh, B., Konopacky, Q.~M., \& Marois, C. 2011, The
  Astrophysical Journal, 733, 65

\bibitem[{Bate(2009)}]{Bate:2009br}
Bate, M.~R. 2009, Monthly Notices RAS, 392, 590

\bibitem[{Boss(1997)}]{Boss:1997di}
Boss, A.~P. 1997, Science, 276, 1836

\bibitem[{Bowler {et~al.}(2014)Bowler, Liu, Kraus, \& Mann}]{Bowler:2014dk}
Bowler, B.~P., Liu, M.~C., Kraus, A.~L., \& Mann, A.~W. 2014, ApJ, 784, 65

\bibitem[{Brandt {et~al.}(2014)Brandt, McElwain, Turner, Mede, Spiegel,
  Kuzuhara, Schlieder, Wisniewski, Abe, Biller, Brandner, Carson, Currie,
  Egner, Feldt, Golota, Goto, Grady, Guyon, Hashimoto, Hayano, Hayashi,
  Hayashi, Henning, Hodapp, Inutsuka, Ishii, Iye, Janson, Kandori, Knapp, Kudo,
  Kusakabe, Kwon, Matsuo, Miyama, Morino, Moro-Martin, Nishimura, Pyo, Serabyn,
  Suto, Suzuki, Takami, Takato, Terada, Thalmann, Tomono, Watanabe, Yamada,
  Takami, Usuda, \& Tamura}]{Brandt:2014cw}
Brandt, T.~D., McElwain, M.~W., Turner, E.~L., {et~al.} 2014, ApJ, 794, 159

\bibitem[{Bulger {et~al.}(2014)Bulger, Patience, Ward-Duong, Pinte, Bouy,
  Menard, \& Monin}]{Bulger:2014ei}
Bulger, J., Patience, J., Ward-Duong, K., {et~al.} 2014, A{\&}A, 570, A29

\bibitem[{Burgasser(2014)}]{Burgasser:2014tr}
Burgasser, A.~J. 2014, arXiv, 1406.4887v1

\bibitem[{Burrows {et~al.}(1997)Burrows, Marley, Hubbard, Lunine, Guillot,
  Saumon, Freedman, Sudarsky, \& Sharp}]{Burrows:1997jq}
Burrows, A., Marley, M., Hubbard, W.~B., {et~al.} 1997, Astrophysical Journal,
  491, 856

\bibitem[{Caffau {et~al.}(2010)Caffau, Ludwig, Steffen, Freytag, \&
  Bonifacio}]{Caffau:2010ik}
Caffau, E., Ludwig, H.~G., Steffen, M., Freytag, B., \& Bonifacio, P. 2010, Sol
  Phys, 268, 255

\bibitem[{Chabrier(2003)}]{Chabrier:2003ki}
Chabrier, G. 2003, The Publications of the Astronomical Society of the Pacific,
  115, 763

\bibitem[{Cushing {et~al.}(2004)Cushing, Vacca, \& Rayner}]{Cushing:2004bq}
Cushing, M.~C., Vacca, W.~D., \& Rayner, J.~T. 2004, PASP, 116, 362

\bibitem[{Davies {et~al.}(2013)Davies, Adams, Armitage, Chambers, Ford,
  Morbidelli, Raymond, \& Veras}]{Davies:2013vb}
Davies, M.~B., Adams, F.~C., Armitage, P., {et~al.} 2013, arXiv, 1311.6816v1

\bibitem[{Dupuy \& Liu(2012)}]{Dupuy:2012bp}
Dupuy, T.~J., \& Liu, M.~C. 2012, The Astrophysical Journal Supplement, 201, 19

\bibitem[{Herczeg \& Hillenbrand(2014)}]{Herczeg:2014is}
Herczeg, G.~J., \& Hillenbrand, L.~A. 2014, ApJ, 786, 97

\bibitem[{Kraus {et~al.}(2014)Kraus, Ireland, cieza, hinkley, Dupuy, Bowler, \&
  Liu}]{Kraus:2014tl}
Kraus, A.~L., Ireland, M.~J., cieza, L.~A., {et~al.} 2014, ApJ, 781, 20

\bibitem[{Kraus {et~al.}(2011)Kraus, Ireland, Martinache, \&
  Hillenbrand}]{Kraus:2011hv}
Kraus, A.~L., Ireland, M.~J., Martinache, F., \& Hillenbrand, L.~A. 2011, The
  Astrophysical Journal, 731, 8

\bibitem[{Lagrange {et~al.}(2010)Lagrange, Bonnefoy, Chauvin, Apai, Ehrenreich,
  Boccaletti, Gratadour, Rouan, Mouillet, Lacour, \& Kasper}]{Lagrange:2010fsa}
Lagrange, A.-M., Bonnefoy, M., Chauvin, G., {et~al.} 2010, Science, 329, 57

\bibitem[{Lambrechts \& Johansen(2012)}]{Lambrechts:2012gr}
Lambrechts, M., \& Johansen, A. 2012, A{\&}A, 544, A32

\bibitem[{Larkin {et~al.}(2006)Larkin, Barczys, Krabbe, Adkins, Aliado, Amico,
  Brims, Campbell, Canfield, Gasaway, Honey, Iserlohe, Johnson, Kress,
  LaFreniere, Lyke, Magnone, Magnone, McElwain, Moon, Quirrenbach, Skulason,
  Song, Spencer, Weiss, \& Wright}]{Larkin:2006jd}
Larkin, J., Barczys, M., Krabbe, A., {et~al.} 2006, Proc. SPIE, 6269, 42

\bibitem[{Li {et~al.}(2015)Li, Kouwenhoven, Stamatellos, \&
  Goodwin}]{Li:2015ib}
Li, Y., Kouwenhoven, M. B.~N., Stamatellos, D., \& Goodwin, S.~P. 2015, ApJ,
  805, 1

\bibitem[{Liu {et~al.}(2010)Liu, Dupuy, \& Leggett}]{Liu:2010cw}
Liu, M.~C., Dupuy, T.~J., \& Leggett, S.~K. 2010, The Astrophysical Journal,
  722, 311

\bibitem[{Low \& Lynden-Bell(1976)}]{Low:1976wt}
Low, C., \& Lynden-Bell, D. 1976, Royal Astronomical Society, 176, 367

\bibitem[{Luhman {et~al.}(2009)Luhman, Mamajek, Allen, \& Cruz}]{Luhman:2009cn}
Luhman, K.~L., Mamajek, E.~E., Allen, P.~R., \& Cruz, K.~L. 2009, The
  Astrophysical Journal, 703, 399

\bibitem[{Markwardt(2009)}]{Markwardt:2009wq}
Markwardt, C.~B. 2009, Astronomical Data Analysis Software and Systems XVIII
  ASP Conference Series, 411, 251

\bibitem[{Marley {et~al.}(2012)Marley, Saumon, Cushing, Ackerman, Fortney, \&
  Freedman}]{Marley:2012fo}
Marley, M.~S., Saumon, D., Cushing, M., {et~al.} 2012, The Astrophysical
  Journal, 754, 135

\bibitem[{Marois {et~al.}(2008)Marois, Macintosh, Barman, Zuckerman, Song,
  Patience, Lafreniere, \& Doyon}]{Marois:2008ei}
Marois, C., Macintosh, B., Barman, T., {et~al.} 2008, Science, 322, 1348

\bibitem[{Marois {et~al.}(2010)Marois, Zuckerman, Konopacky, Macintosh, \&
  Barman}]{Marois:2010gpa}
Marois, C., Zuckerman, B., Konopacky, Q.~M., Macintosh, B., \& Barman, T. 2010,
  Nature, 468, 1080

\bibitem[{Mieda {et~al.}(2014)Mieda, Wright, Larkin, Graham, Adkins, Lyke,
  Campbell, Maire, Do, \& Gordon}]{Mieda:2014dt}
Mieda, E., Wright, S.~A., Larkin, J.~E., {et~al.} 2014, PASP, 126, 250

\bibitem[{Natta {et~al.}(2004)Natta, Testi, Muzerolle, Randich, Comer~N, \&
  Persi}]{Natta:2004kz}
Natta, A., Testi, L., Muzerolle, J., {et~al.} 2004, A{\&}A, 424, 603

\bibitem[{Padoan \& Nordlund(2004)}]{Padoan:2004ez}
Padoan, P., \& Nordlund, {\AA}. 2004, The Astrophysical Journal, 617, 559

\bibitem[{Pineda {et~al.}(2015)Pineda, Offner, Parker, Arce, Goodman, Caselli,
  Fuller, Bourke, \& Corder}]{Pineda:2015iz}
Pineda, J.~E., Offner, S. S.~R., Parker, R.~J., {et~al.} 2015, Nature, 518, 213

\bibitem[{Pollack {et~al.}(1996)Pollack, Hubickyj, Bodenheimer, Lissauer,
  Podolak, \& Greenzweig}]{Pollack:1996jp}
Pollack, J.~B., Hubickyj, O., Bodenheimer, P., {et~al.} 1996, Icarus, 124, 62

\bibitem[{Rayner {et~al.}(2009)Rayner, Cushing, \& Vacca}]{Rayner:2009ki}
Rayner, J.~T., Cushing, M.~C., \& Vacca, W.~D. 2009, The Astrophysical Journal
  Supplement, 185, 289

\bibitem[{Reipurth \& Mikkola(2015)}]{Reipurth:2015dz}
Reipurth, B., \& Mikkola, S. 2015, The Astronomical Journal, 149, 1

\bibitem[{Stamatellos \& Whitworth(2009)}]{Stamatellos:2009fw}
Stamatellos, D., \& Whitworth, A.~P. 2009, Monthly Notices RAS, 392, 413

\bibitem[{Todorov {et~al.}(2010)Todorov, Luhman, \& Mcleod}]{Todorov:2010cn}
Todorov, K., Luhman, K.~L., \& Mcleod, K.~K. 2010, The Astrophysical Journal,
  714, L84

\bibitem[{Todorov {et~al.}(2014)Todorov, Luhman, Konopacky, Mcleod, Apai, Ghez,
  Pascucci, \& Robberto}]{Todorov:2014fq}
Todorov, K.~O., Luhman, K.~L., Konopacky, Q.~M., {et~al.} 2014, ApJ, 788, 40

\bibitem[{Tokovinin(2014)}]{Tokovinin:2014ea}
Tokovinin, A. 2014, The Astronomical Journal, 147, 87

\bibitem[{Vacca {et~al.}(2003)Vacca, Cushing, \& Rayner}]{Vacca:2003wi}
Vacca, W.~D., Cushing, M.~C., \& Rayner, J.~T. 2003, PASP, 115, 389

\bibitem[{van Dam {et~al.}(2006)van Dam, Bouchez, Le~Mignant, Johansson,
  Wizinowich, Campbell, Chin, Hartman, Lafon, Stomski, \&
  Summers}]{vanDam:2006hw}
van Dam, M.~A., Bouchez, A.~H., Le~Mignant, D., {et~al.} 2006, Publications of
  the Astronomical Society of the Pacific, 118, 310

\bibitem[{White {et~al.}(1999)White, Ghez, Reid, \& Schultz}]{White:1999bt}
White, R.~J., Ghez, A.~M., Reid, I.~N., \& Schultz, G. 1999, The Astrophysical
  Journal, 520, 811

\bibitem[{Whitworth \& Stamatellos(2006)}]{Whitworth:2006cd}
Whitworth, A.~P., \& Stamatellos, D. 2006, A{\&}A, 458, 817

\bibitem[{Wizinowich {et~al.}(2006)Wizinowich, Le~Mignant, Bouchez, Campbell,
  Chin, Contos, van Dam, Hartman, Johansson, Lafon, Lewis, Stomski, Summers,
  Brown, Danforth, Max, \& Pennington}]{Wizinowich:2006hk}
Wizinowich, P.~L., Le~Mignant, D., Bouchez, A.~H., {et~al.} 2006, The
  Publications of the Astronomical Society of the Pacific, 118, 297

\bibitem[{Wu {et~al.}(2015)Wu, Close, Males, Barman, Morzinski, Follette,
  Bailey, Rodigas, Hinz, Puglisi, Xompero, \& Briguglio}]{Wu:2015kh}
Wu, Y.-L., Close, L.~M., Males, J.~R., {et~al.} 2015, ApJ, 801, 4

\bibitem[{Yelda {et~al.}(2010)Yelda, Lu, Ghez, Clarkson, Anderson, Do, \&
  Matthews}]{Yelda:2010ig}
Yelda, S., Lu, J.~R., Ghez, A.~M., {et~al.} 2010, The Astrophysical Journal,
  725, 331

\end{thebibliography}

\clearpage

\begin{deluxetable}{lcccccccccc}
\tabletypesize{\scriptsize}
\setlength{ \tabcolsep } {.12cm} 
\tablewidth{0pt}
\tablecolumns{11}
\tablecaption{Relative Photometry and Astrometry \label{tab:obs}}
\tablehead{
   \colhead{Name} & \colhead{UT Date} & \colhead{Inst.}  & \colhead{$N$$\times$Exp.}  & \colhead{Filt.}  & \colhead{Sep.}      & \colhead{P.A.}         & \colhead{$\Delta$mag} & \colhead{$m_A$} &  \colhead{$m_B$}  &  \colhead{FWHM} \\
      \colhead{}         & \colhead{}                & \colhead{}          & \colhead{Time (s)}                                                                         & \colhead{}         &  \colhead{($''$)}           & \colhead{ ($^{\circ}$)} &  \colhead{}           &   \colhead{(mag)}     &  \colhead{mag}     & \colhead{(mas)}
        }
\startdata
2M0441Aab  &  2014 Nov 10  &  NIRC2  &  7 $\times$ 15  &  $Y$  &  230 $\pm$ 2  &  79.3 $\pm$ 0.6  &  3.16 $\pm$ 0.16  &  11.37 $\pm$ 0.02  &  14.53 $\pm$ 0.15 &  96 $\pm$ 29\\
2M0441Aab  &  2014 Nov 10  &  NIRC2  &  11 $\times$ 5  &  $J$  &  234 $\pm$ 2  &  79.4 $\pm$ 0.6  &  2.96 $\pm$ 0.06  &  10.77 $\pm$ 0.04  &  13.73 $\pm$ 0.07 & 65 $\pm$ 12\\
2M0441Aab  &  2014 Nov 10  &  NIRC2  &  7 $\times$ 5  &  $H$  &  233.0 $\pm$ 1.6  &  79.9 $\pm$ 0.2  &  3.11 $\pm$ 0.07  &  10.19 $\pm$ 0.03  &  13.30 $\pm$ 0.07 & 65 $\pm$ 17\\
2M0441Aab  &  2014 Nov 10  &  NIRC2  &  6 $\times$ 5  &  $K_S$  &  232.0 $\pm$ 1.3  &  79.1 $\pm$ 0.3  &  2.77 $\pm$ 0.13  &  9.94 $\pm$ 0.02  &  12.71 $\pm$ 0.12 & 64 $\pm$ 5\\
2M0441Aab  &  2014 Dec 08  &  OSIRIS  &  8 $\times$ 300  &  $Jbb$  &  $\cdots$  &  $\cdots$  &  3.3 $\pm$ 0.3  &  10.70 $\pm$ 0.05  &  14.0 $\pm$ 0.3 &  54 $\pm$ 5 \\
2M0441Aab  &  2014 Dec 08  &  OSIRIS  &  6 $\times$ 300  &  $Hbb$  &  $\cdots$  &  $\cdots$  &  3.14 $\pm$ 0.18  &  10.20 $\pm$ 0.03  &  13.34 $\pm$ 0.17 &  58 $\pm$ 3  \\
2M0441Aab  &  2014 Dec 08  &  OSIRIS  &  6 $\times$ 300  &  $Kbb$  &  $\cdots$  &  $\cdots$  &  2.81 $\pm$ 0.10  &  9.96 $\pm$ 0.02  &  12.77 $\pm$ 0.10 &  54 $\pm$ 1 \\
2M0441Bab  &  2014 Nov 10  &  NIRC2  &  7 $\times$ 60  &  $H$  &  93 $\pm$ 2  &  132.6 $\pm$ 1.1  &  1.45 $\pm$ 0.14  &  14.03 $\pm$ 0.05  &  15.48 $\pm$ 0.12 & 88 $\pm$ 10\\
2M0441Bab  &  2014 Nov 10  &  NIRC2  &  7 $\times$ 30  &  $Ks$  &  97.5 $\pm$ 1.0  &  132.0 $\pm$ 0.6  &  1.24 $\pm$ 0.05  &  13.46 $\pm$ 0.03  &  14.70 $\pm$ 0.05 &  65 $\pm$ 4\\
2M0441Bab  &  2014 Dec 08  &  OSIRIS  &  10 $\times$ 400  &  $Jbb$  &  $\cdots$  &  $\cdots$  &  1.7 $\pm$ 0.2  &  14.57 $\pm$ 0.06  &  16.26 $\pm$ 0.17  &  67 $\pm$ 7 \\
2M0441Bab  &  2014 Dec 08  &  OSIRIS  &  8 $\times$ 300  &  $Hbb$  &  $\cdots$  &  $\cdots$  &  1.42 $\pm$ 0.19  &  14.06 $\pm$ 0.05  &  15.47 $\pm$ 0.15  &  88 $\pm$ 9 \\
2M0441Bab  &  2014 Dec 08  &  OSIRIS  &  8 $\times$ 300  &  $Kbb$  &  $\cdots$  &  $\cdots$  &  1.35 $\pm$ 0.17  &  13.50 $\pm$ 0.05  &  14.85 $\pm$ 0.14 &  79 $\pm$ 10 
\enddata
\end{deluxetable}
\clearpage

\clearpage
\newpage

\acknowledgments

\end{document}